\documentclass{aa}
\usepackage[varg]{txfonts}
\usepackage{graphicx}
\usepackage{ulem}
\usepackage{color}

\usepackage{natbib}
\bibpunct{(}{)}{;}{a}{}{,} 

\definecolor{orange}{rgb}{ 0.95, 0.60, 0}

\begin{document}

\title{Formation of the terrestrial planets in the solar system around 1 au via radial concentration of planetesimals}
\titlerunning{Formation of solar system's terrestrial planets without forming close-in planets}
\author{Masahiro Ogihara\inst{\ref{inst1}}
\and Eiichiro Kokubo\inst{\ref{inst1}}
\and Takeru K. Suzuki\inst{\ref{inst2}}
\and Alessandro Morbidelli\inst{\ref{inst3}}
}
\institute{Division of Theoretical Astronomy, National Astronomical Observatory of Japan, 2-21-1, Osawa, Mitaka, 181-8588 Tokyo, Japan \email{masahiro.ogihara@nao.ac.jp}\label{inst1}
\and School of Arts \& Sciences, University of Tokyo, 3-8-1, Komaba, Meguro, 153-8902 Tokyo, Japan\label{inst2}
\and Laboratoire Lagrange, Universit\'e C\^ote d'Azur, Observatoire de la C\^ote d'Azur, CNRS,
Blvd de l'Observatoire, CS 34229, 06304 Nice Cedex 4, France\label{inst3}
}
\authorrunning{M. Ogihara et al.}
\date{Received 17 January 2018 / Accepted 30 March 2018}

\abstract 
{
No planets exist inside the orbit of Mercury and the terrestrial planets of the solar system exhibit a localized configuration. According to thermal structure calculation of protoplanetary disks, a silicate condensation line ($\sim 1300 {\rm ~K}$) is located around 0.1 au from the Sun except for the early phase of disk evolution, and planetesimals could have formed inside the orbit of Mercury.
A recent study of disk evolution that includes magnetically driven disk winds showed that the gas disk obtains a positive surface density slope inside $\sim 1~{\rm au}$ from the central star. In a region with positive midplane pressure gradient, planetesimals undergo outward radial drift.
} 
{
We investigate the radial drift of planetesimals and type I migration of planetary embryos in a disk that viscously evolves with magnetically driven disk winds. We show a case in which no planets remain in the close-in region.
} 
{
Radial drifts of planetesimals are simulated using a recent disk evolution model that includes effects of disk winds. The late stage of planet formation is also examined by performing \textit{N}-body simulations of planetary embryos.
} 
{
We demonstrate that in the middle stage of disk evolution, planetesimals can undergo convergent radial drift in a magnetorotational instability (MRI)-inactive disk, in which the pressure maximum is created, and accumulate in a narrow ring-like region with an inner edge at $\sim$ 0.7~au from the Sun. We also show that planetary embryos that may grow from the narrow planetesimal ring do not exhibit significant type I migration in the late stage of disk evolution.
} 
{
The origin of the localized configuration of the terrestrial planets of the solar system, in particular the deficit of close-in planets, can be explained by the convergent radial drift of planetesimals in disks with a positive pressure gradient in the close-in region.
}
\maketitle

\section{Introduction}

The properties of the terrestrial planets of our solar system are well understood compared to other systems. A number of \textit{N}-body simulations of planetary accretion were performed to reproduce these properties, such as small eccentricities and late moon formation (e.g., \citealt{nagasawa_etal05}; \citealt{ogihara_etal07}; \citealt{raymond_etal09}; \citealt{morishima_etal10}).

One of the major issues facing the theory of planet formation is reproducing the current orbital configuration of the solar system's terrestrial planets. Their orbital locations are confined to a region between 0.38 and 1.5 au from the Sun. Several models have been proposed to reproduce the outer boundary of the localized distribution which includes the grand tack model (e.g., \citealt{walsh_etal11}). In the meantime, it is quite hard to explain why there is no large planets in close-in orbits. If planetesimals form when the condensation line for silicates is located inside $r = 0.7 {\rm ~au}$, protoplanets/planets should form in close-in orbits. However, no planets exist inside the orbit of Mercury in our solar system. Regarding this problem, \citet{ogihara_etal15} demonstrated that protoplanets larger than Mars that form in close-in orbits can undergo outward type I migration in disks with positive surface density profiles, which would help to reproduce the deficit of close-in planets.

Based on three-dimensional (3D) magnetohydrodynamic (MHD) simulations, \citet{suzuki_etal16} derived the long-term global disk evolution including effects of magnetically driven disk winds. They considered the mass loss due to disk winds \citep{suzuki_inutsuka09} and the wind-driven accretion (\citealt{bai_stone13}; \citealt{gressel_etal15}; magnetic braking) in addition to the standard viscous evolution. They found that the disk profile in the close-in region ($r \lesssim 1 {\rm ~au}$) can be significantly altered from the minimum-mass solar nebular (MMSN) model (\citealt{weidenschilling77}; \citealt{hayashi81}). In particular, the slope of the gas surface density is almost flat inside $r \simeq 1 {\rm ~au}$ for MRI-active disk, and the slope can even be positive for MRI-inactive disk. In a separate paper (\citealt{ogihara_etal18b}, hereafter Paper~II), we investigated the orbital evolution of planetary embryos larger than Mars in such disks under various conditions, and found that type I migration of protoplanets can be significantly suppressed in many cases compared to the type I migration rate estimated for the MMSN model. On the other hand, it is expected that smaller particles (dust/planetesimals) can undergo significant radial drift in MRI-inactive disks of \citet{suzuki_etal16}.

In this work, we examine another path to form the localized configuration of the solar system's terrestrial planets using the recent disk evolution model of \citet{suzuki_etal16}. We examine whether or not the following formation scenario can take place by numerical simulations: Planetesimals undergo convergent radial drift to form localized planetesimal disks, then planetary embryos grow to terrestrial planets without undergoing significant type I migration.

The paper is organized as follows. In Section~\ref{sec:model}, we present the numerical model of disk evolution. In Section~\ref{sec:radial_drift}, we show the outcomes of simulations that consider the radial drift of planetesimals. In Section~\ref{sec:n-body}, we present results of \textit{N}-body simulations from planetary embryos. In Section~\ref{sec:conclusions}, we make concluding remarks.

\section{Disk model}
\label{sec:model}

For evolution of disk surface density, we numerically solve the diffusion equation that includes effects of disk winds based on MHD simulations (the same as used by \citealt{suzuki_etal16}; see also \citealt{simon_etal15}; \citealt{hasegawa_etal17}):
\begin{eqnarray}
\label{eq:diffusion}
\frac{\partial \Sigma_{\rm g}}{\partial t} = \frac{1}{r} \frac{\partial}{\partial r} \left[\frac{2}{r\Omega} \left\{ \frac{\partial}{\partial r} (r^2 \Sigma_{\rm g} \overline{\alpha_{r,\phi}} c_{\rm s}^2) + r^2 \overline{\alpha_{\phi,z}} \frac{\Sigma_{\rm g} H \Omega^2}{2 \sqrt{\pi}} \right\} \right] \nonumber \\
 - C_{\rm w} \frac{\Sigma_{\rm g} \Omega}{\sqrt{2 \pi}},
\end{eqnarray}
where $\Sigma_{\rm g}$, $r$, $\Omega$, $c_{\rm s}$, and $H$ are the gas surface density, the radial distance, the Keplerian frequency, the sound speed, and the disk scale height, respectively. An effective turbulent viscosity $\overline{\alpha_{r,\phi}}$ is based on a \citet{shakura_sunyaev73} $\alpha$ viscosity parameterization. Parameters $\overline{\alpha_{\phi,z}}$ and $C_{\rm w}$ describe measures of the angular momentum loss due to the wind torque and mass loss due to disk winds, respectively. According to results of MHD simulations, we chose the parameters as follows.  We adopt two values of the effective turbulent viscosity; namely, $\overline{\alpha_{r,\phi}} = 8 \times 10^{-3}$ (MRI-active case) and $\overline{\alpha_{r,\phi}} = 8 \times 10^{-5}$ (MRI-inactive case). Other parameters are fixed as $\overline{\alpha_{\phi,z}} = 10^{-5} (\Sigma_{\rm g} / \Sigma_{\rm g,ini})^{-0.66}$ \citep{bai13}, $C_{\rm w} = 2 \times 10^{-5}$ (MRI-active case; \citealt{suzuki_inutsuka09}) and $C_{\rm w} = 10^{-5}$ (MRI-inactive case; \citealt{suzuki_etal10}). As described in Section~2.3 of \citet{suzuki_etal16}, the mass-loss rate due to disk winds $C_{\rm w}$ is also constrained by energetics.

The initial gas profile is given by $\Sigma_{\rm g,ini} = 1.7 \times 10^4 (r/1 {\rm ~au})^{-3/2} \exp(-r/30 {\rm ~au}) ~{\rm g~cm^{-2}}$.
In order to consider the early stage of disk evolution, which is connected to star formation, the initial gas disk is about ten times more massive than the MMSN model. We note that planetesimals and planetary embryos take some time to form, therefore we start simulations some time after the disk formation. In the same way as in our previous works, we start simulations at $t=10^5 {\rm ~yr}$ in the disk evolution of \citet{suzuki_etal16}, and refer to this time as ``initial.''

\begin{figure}
\resizebox{1.0 \hsize}{!}{\includegraphics{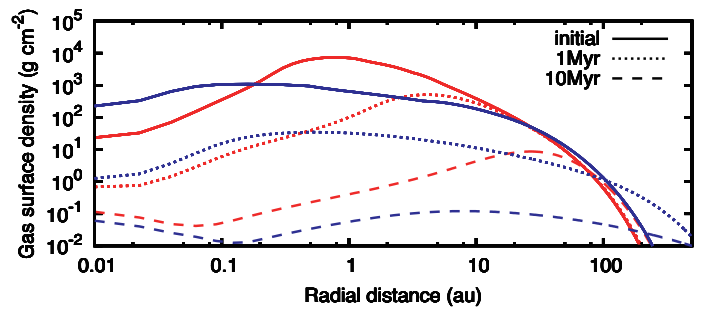}}
\caption{Time evolution of gas surface density based on \citet{suzuki_etal16}. Blue and red lines represent MRI-active disk and MRI-inactive disk, respectively.
}
\label{fig:r_sigma}
\end{figure}

Figure~\ref{fig:r_sigma} shows the time evolution of the gas surface density for MRI-active disks and MRI-inactive disks. The evolution is the same as in \citet{suzuki_etal16}. We see that disk profiles are altered in the close-in region $(r \lesssim 1 {\rm ~au})$. An important point is that the disk obtains a positive surface density slope in a broad close-in region in MRI-inactive disks. As discussed in \citet{suzuki_etal16}, the angular momentum loss due to the wind torque is inversely correlated with the gas density. In addition, the mass loss rate is proportional to the Keplerian time. Therefore, the gas density is carved out from inside based on the model of \citet{suzuki_etal16}. It is likely that in the late stage of disk evolution, the disk is MRI active due to smaller gas surface density. For the evolution of the disk temperature, the stellar irradiation and viscous heating are considered. We refer to Section~2.4 of \citet{suzuki_etal16} for a detailed description.

\section{Radial drift of planetesimals}
\label{sec:radial_drift}

We first examine the radial drift of planetesimals before the formation of planetary embryos. Particles with the size of planetesimals or smaller undergo radial migration due to gas drag. As stated in Section~\ref{sec:model}, it is considered that the MRI activity increases with decreasing gas surface density. Therefore, here we  use the MRI-inactive disk for the orbital evolution phase of planetesimals. 
As discussed in \citet{suzuki_etal16}, disk winds affect the radial drift of planetesimals. As the radial pressure gradient force and hence the rotation velocity of gas are altered, planetesimals can feel a tail wind in the region of a positive pressure gradient, leading to outward drift.

\begin{figure}
\begin{center}
\resizebox{0.7 \hsize}{!}{\includegraphics{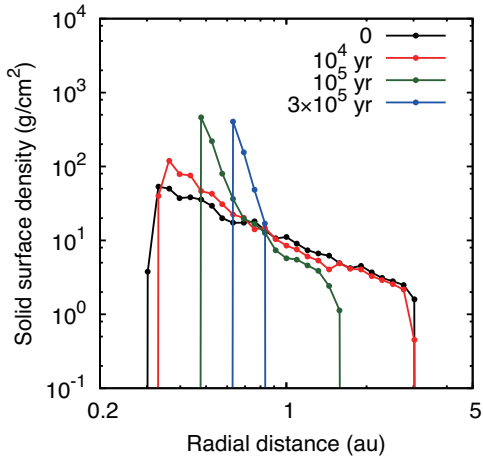}}
\end{center}
\caption{Time evolution of solid surface density. Planetesimals with $R \simeq 1 {\rm ~km}$ are placed between $r = 0.3$ and 3~au. The disk is considered to be MRI-inactive.
}
\label{fig:a_sigma}
\end{figure}

Here we perform simulations of the orbital evolution of planetesimals in MRI-inactive disks with mass $M = 1 \times 10^{16} ~{\rm g}$ ($R \simeq 1 {\rm ~km}$ assuming the internal density of 3 $~{\rm g~cm^{-3}}$) that are placed between $r = 0.3$~au and 3~au. Mutual interactions between planetesimals (i.e., gravity, collision) are ignored. For aerodynamical gas drag force, we use the formula of \citet{adachi_etal76}. The drift timescale (see Appendix~\ref{sec:app1}) is inversely proportional to a scaling factor of pressure gradient force $\eta = -1/2 (c_{\rm s}/v_{\rm K})^2 (d \ln P/d \ln r)$.
See Figure~\ref{fig:app} for the evolution of $\eta$. Figure~\ref{fig:a_sigma} shows the result of our simulation, in which the evolution of solid surface density is indicated. We find that planetesimals undergo convergent radial drift. As the disk evolves, the location of the pressure maximum moves outward. After $t = 0.3 ~{\rm Myr}$, planetesimals accumulate in a narrow ring-like region because all planetesimals inside $\sim 0.7~{\rm au}$ undergo outward drift. Therefore, it is possible that a localized configuration is attained by radial drift of planetesimals before formation of planetary embryos. 
We note that typical sizes of planetesimals that form via streaming instability would be larger than 1~km (e.g., \citealt{simon_etal16}). Larger planetesimals suffer less efficient radial drift. Thus, if planetesimals larger than 10~km are distributed between 0.3 and 3~au, the strong convergent drift seen in Fig.~\ref{fig:a_sigma} does not occur.

Although it is reasonable to assume that the disk is MRI inactive in the early/middle phase of planet formation, we also comment on radial drift of planetesimals in MRI active disks.
When the surface density distribution exhibits flat profiles (e.g., MRI-active disk), planetesimals undergo slow inward drift instead of outward drift. In this case, narrow ring-like regions would not form; however, as the drift velocity is slowed down significantly, it would be useful to overcome the drift barrier of dust/planetesimals.

We note that we adopt several simplifications that  affect the realism of the simulation. First,  a single size population of planetesimals is considered, and their growth is also ignored. In reality, planetesimals have size distribution, which changes with time. Second, mutual interactions between planetesimals and the back reaction of planetesimals on the gas disk are neglected. When solid particles accumulate in a narrow ring, where the dust-to-gas ratio is close to unity, the back reaction on the gas disk should be important. The evolution in Fig.~\ref{fig:a_sigma} is therefore an idealized result. We will investigate these effects in future works.

\section{Orbital evolution of planetary embryos via type I migration}
\label{sec:n-body}

Now we examine the late stage of planet formation from planetary embryos. As we showed in  the
previous subsection, planetesimals can accumulate in a narrow-ring region. We therefore  consider the subsequent evolution for the initial embryo distribution. \citet{hansen09} adopted a localized configuration for the initial embryo distribution and performed \textit{N}-body simulations. We use the same initial conditions as \citet{hansen09}.
As an initial condition of \textit{N}-body simulations, 400 planetary embryos with mass $M = 0.005 ~M_\oplus$ are randomly placed between 0.7 and 1 au according to a uniform surface density. Their initial eccentricities and inclinations are set to be small ($\simeq 10^{-2}$). Although \citet{hansen09} considered the presence of Jupiter, we do not include any giant planets in our simulations.

For the disk evolution, as we stated in the previous section, it is reasonable to consider that the disk is MRI-inactive in the early phase, which evolves to the MRI-active disk. However, the MRI activity is uncertain when embryos form; therefore we perform \textit{N}-body simulations of embryos both in MRI-active and -inactive disks. The disk evolution model is already shown in Section~\ref{sec:model}.

For the formulae of type I migration, we use those described in \citet{paardekooper_etal11}. The unsaturated torque is given by $\Gamma = (-0.6 + 2.3p - 2.8q) \Gamma_0,$ where $p = \mathrm{d}\ln \Sigma_{\rm g}/ \mathrm{d}\ln r$ is the local surface density slope and $q= \mathrm{d}\ln T/ \mathrm{d}\ln r$ is the local temperature slope \citep{paardekooper_etal10}. In our disk model, the temperature slope is roughly given by $q = -1/2$ (see Fig.~5 of \citealt{suzuki_etal16}), therefore the rate and direction of type I migration is mainly determined by the surface density slope $q$. The actual torque is calculated by considering the saturation effect. Here the saturation is determined by the viscous diffusion coefficient $\nu$, which is proportional to $\overline{\alpha_{r,\phi}}$ (see Section 4.1 of \citealt{ogihara_etal17}).

\begin{figure}
\resizebox{1.0 \hsize}{!}{\includegraphics{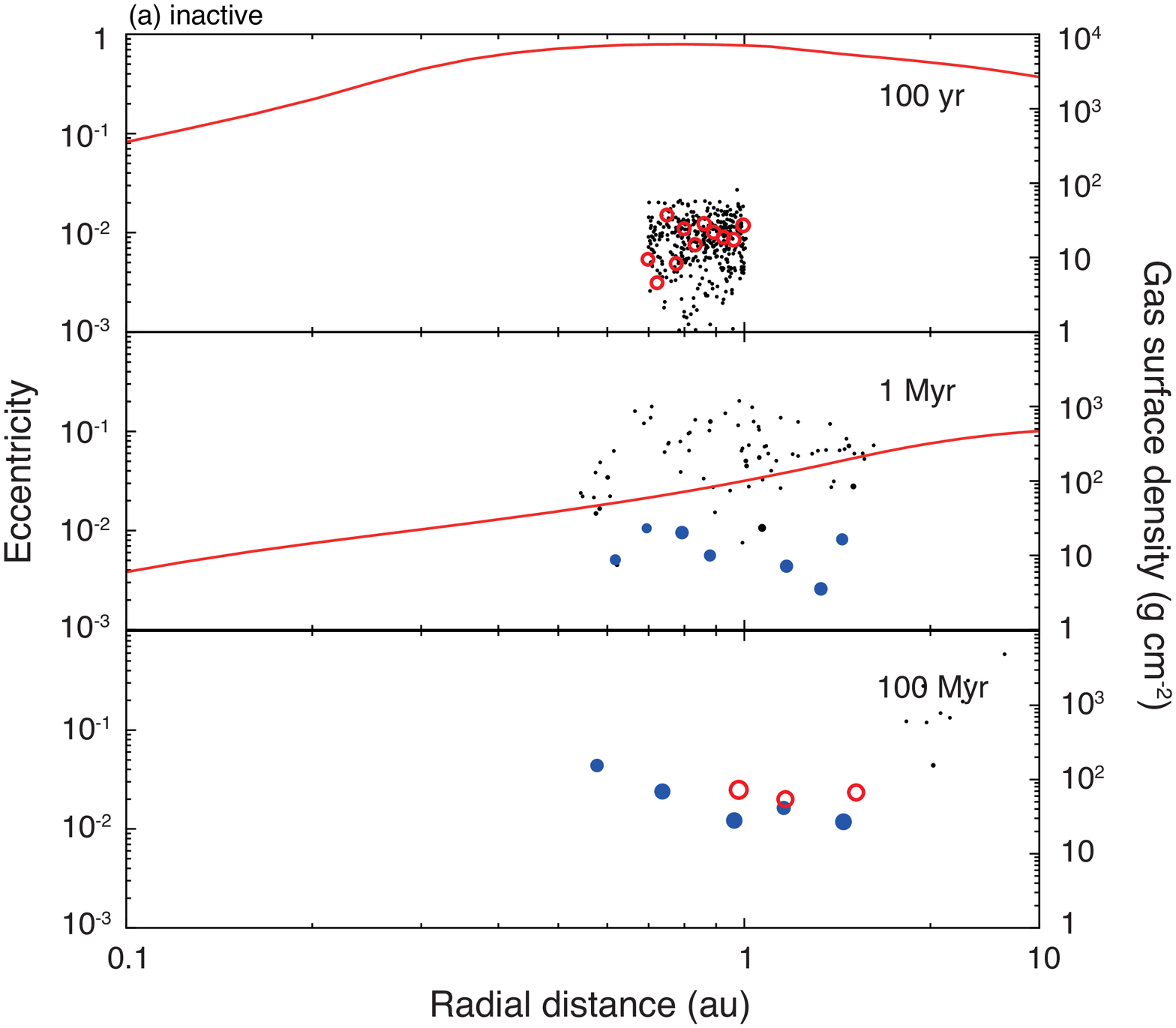}}\\
\resizebox{1.0 \hsize}{!}{\includegraphics{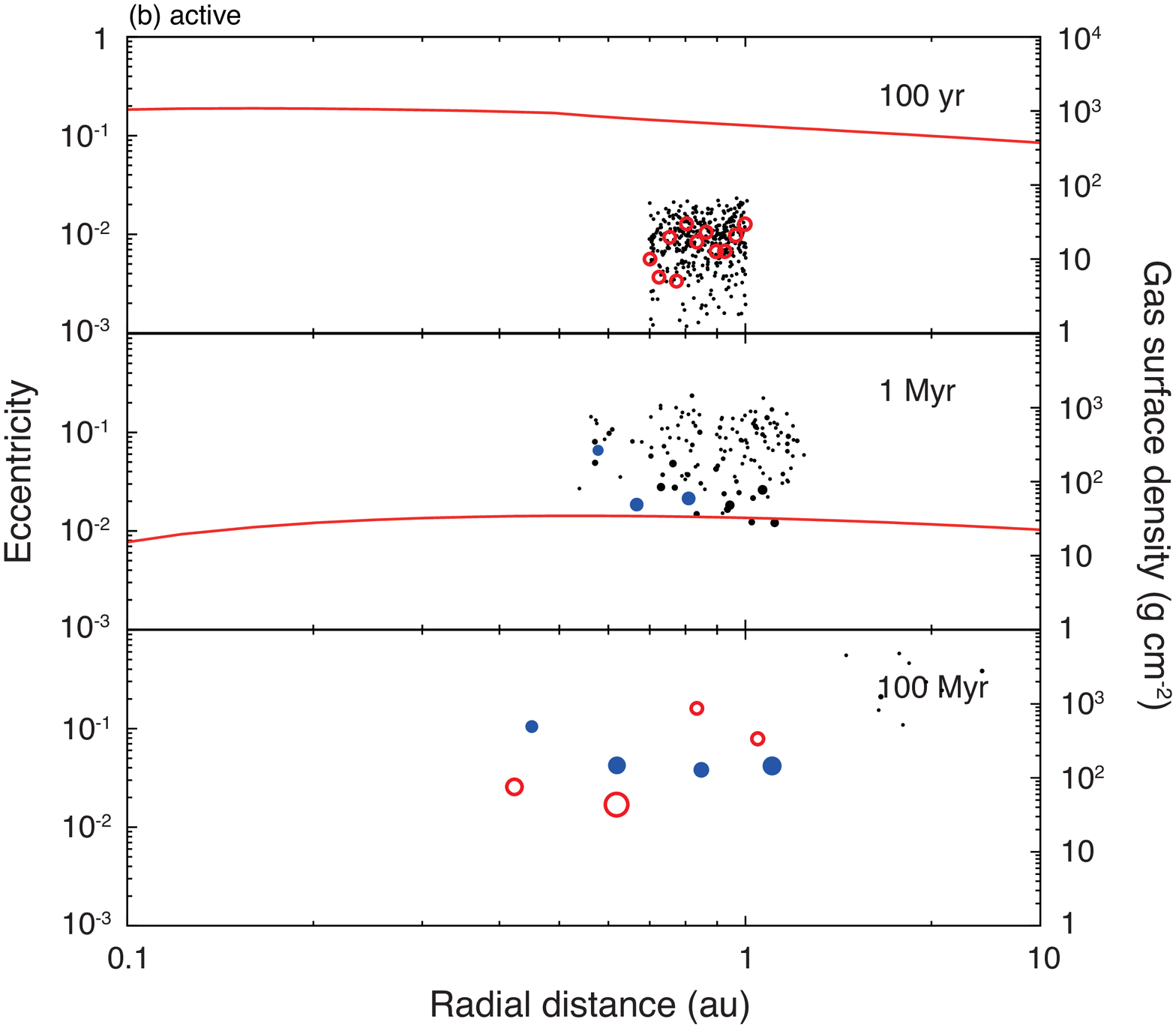}}
\caption{Snapshots of the system for MRI-inactive disk (a) and MRI-active disk (b). The filled circles represent the size of particles. When particles grow to more massive than $0.1~M_\oplus$, they are indicated by filled blue circles. The red lines indicate the gas surface density (right axis). The red circles represent the initial and final orbital configurations of simulations from isolation-mass protoplanets.
}
\label{fig:t_a_ss}
\end{figure}

Figure~\ref{fig:t_a_ss}(a) shows snapshots of the system for MRI inactive disks (see also Paper~II for the migration map). No significant inward migration is seen. Although some protoplanets move outward to $r \simeq 2 {\rm ~au}$, the innermost larger protoplanets ($M \gtrsim 0.1~M_\oplus$) are located at $r = 0.56 {\rm ~au}$ at $t = 100 {\rm ~Myr}$. 

Figure~\ref{fig:t_a_ss}(b) shows results of the \textit{N}-body simulation for MRI-active disks. We find no clear tendency for planets to migrate inward or outward. In the final state ($t = 100 {\rm ~Myr}$), four larger protoplanets remain between $r = 0.46 {\rm ~au}$ and 1.1 au. The region where planets exist spreads compared to the initial ring-like region. This is not because planets exhibit migration but because they undergo scattering between protoplanets. 

For comparison, we perform \textit{N}-body simulations for the MMSN model. We find that after planetary embryos grow to Mars-mass protoplanets ($M \simeq 0.1~M_\oplus$), they rapidly migrate inward and are lost to the central star with a migration timescale of $\simeq 0.1~{\rm Myr}$. At $t = 10 {\rm ~Myr}$, the total mass of remaining objects is only $0.445~M_\oplus$. Therefore we can confirm that disk winds play a role in suppressing type I migration, which leads to localized orbital configurations.  We note that from geochemical constraints, Mars grew to half of its present size in about 1.8 Myr \citep{dauphas_pourmand11}. If Mars grew on this timescale and the MMSN decays exponentially with a timescale of 1 Myr, Mars would migrate marginally and not fall onto the star.

We have also performed additional simulations that start from protoplanets with an isolation mass instead of embryos with $M = 0.005~M_\oplus$. For the calculation of the isolation mass, see Eq.~(13) of \citet{kokubo_ida02} for example. We chose a relatively small initial orbital separation of $5~r_{\rm H}$, where $r_{\rm H}$ represents the mutual Hill radius of neighboring protoplanets. The initial and final orbital configuration is overplotted in Figure~\ref{fig:t_a_ss} by circles. We find that our conclusion does not change. No planets remain in the close-in region ($a < 0.4 {\rm ~au}$) and planetary orbits are confined to a localized region. Detailed results of \textit{N}-body simulations of isolation-mass protoplanets are presented in Paper~II.

\section{Conclusions}
\label{sec:conclusions}

We investigate the formation of the solar system's terrestrial planets in disks evolving with disk winds. We demonstrate that planetesimals formed in close-in orbits can undergo outward radial drift in MRI-inactive disks and the inner edge of the planetesimal ring is created at $r \sim $0.7~au. Subsequently, we examine the orbital evolution and growth of embryos from a narrow ring by \textit{N}-body simulation, and find that migration of protoplanets can be significantly suppressed. As a result, we confirm that no large planets remain inside $\simeq 0.5  {\rm ~au}$. \citet{hansen09} showed that the current orbital configuration of the solar system's terrestrial planets can be reproduced if a narrow solid ring is formed in a gas-free environment. Here, in this work, even if we start simulations from an earlier stage when gas is present, we show that the localized configuration can be maintained. 
In a previous work, \citet{ogihara_etal15} showed that protoplanets can undergo outward migration to $\simeq 1 {\rm ~au}$ in a region with a positive surface density slope. In the present work, we showed that even if planetary embryos do not undergo convergent migration, the localized configuration can be created due to convergent radial drift of planetesimals.
The reason why all planetesimals that  formed in the close-in region undergo outward radial drift, which creates the inner edge of planetesimal ring, is that the gas disk obtains a positive pressure gradient in a broad close-in region, which expands outwards in time. 
This results in outward drift of all planetesimals in the close-in region, leaving no planetesimals distributed in the broad close-in region.
This is qualitatively different from other narrower pressure traps (e.g., the snow line).
Such traps can stop inward drift, but cannot drag planetesimals outwards over the full close-in region. If there was no clear disk inner edge, the narrow-trap case would give a ring-like planetesimal distribution because planetesimals inside the pressure trap would fall onto the star. However, if there existed an inner edge, inward drifting planetesimals would be accumulated there; the planetesimal distribution could be very different from our model.
We note also that the outer boundary is determined by the initial distribution of planetesimals/embryos.  Other models are required (e.g.,
the grand tack model) to explain the origin of the outer boundary in the solar system's terrestrial planets.

In Section~\ref{sec:radial_drift}, planetesimals are initially distributed between 0.3 and 3 au. Although we do not specify their origin, they would be formed via radial drift of dust and pebbles, which undergo rapid drift. We note here that no pressure bump exists in the early phase of disk evolution of \citet{suzuki_etal16}, and it takes some time ($\sim$ 0.1 Myr) to form positive slope in the MRI-inactive disk. Therefore, in the earlier stage, pebbles can drift inwards with no limit and planetesimals can form everywhere in the close-in region, which justifies our choice of initial planetesimal distribution.
On the other hand, if a pressure bump exists before the planetesimal formation, dust can accumulate at the pressure maximum and planetesimals only form in a local region at the pressure bump. In this case, the planetesimal distribution would differ from that used in Section~\ref{sec:radial_drift}. However, our conclusion would still be valid because this situation is consistent with our assumption in Section~\ref{sec:n-body} that embryos form in a narrow ring-like region. In this case, the radial drift of planetesimals is not needed and big ($>$ 10~km) planetesimals could form.
In either case, as we stated in Section~\ref{sec:radial_drift},  a simulation of growth and radial drift of dust particles in a consistent manner is required. We note also that an alternative mechanism to create the narrow planetesimal ring by drifting pebbles and their backreaction to the disk has been proposed \citep{drazkowska_etal16}.

In addition, disk winds would contribute to a solution of the ``snow line problem.'' According to radiative transfer calculations, the ice condensation line (snow line) at the midplane moves inside 1~au when the accretion rate of the disk drops below $\simeq 10^{-9}~M_\odot / {\rm yr}$ (e.g., \citealt{oka_etal11}). If planetesimals form at this time and the Earth forms in situ, the Earth should accrete a large amount of icy material, and it is hard to explain the origin of the amount of water (0.023 wt\%) on the Earth. 
Here we apply the disk-wind model to this problem. When the snow line moves inside 1~au, we can assume that the pressure bump is located outside 1~au. If planetary embryos or planetesimals that grow to form the Earth were to form at 1~au by the time of the passage of the snow line through 1~au, the pressure maximum outside 1~au would possibly prevent the inward drift of icy pebbles in the vicinity of the Earth. This could be another mechanism of a fossilization of the snow line proposed by \citet{morbidelli_etal16}. This point should be explored in future works.

\begin{acknowledgements}
We thank Cornelis Dullemond for constructive comments.
Numerical computations were conducted on the general-purpose PC farm at the Center for Computational Astrophysics, CfCA, of the National Astronomical Observatory of Japan. This work was supported by JSPS KAKENHI Grant Numbers 16H07415 and 17H01105.
\end{acknowledgements}

\appendix

{}


\begin{thebibliography}{}

\bibitem[Adachi et al.(1976)]{adachi_etal76}
Adachi, I., Hayashi, C., \& Nakazawa, K. 1976,
Prog. Theor. Phys., 56, 1756
\bibitem[Bai(2013)]{bai13}
Bai, X.-N. 2013,
\apj, 772, 96
\bibitem[Bai \& Stone(2013)]{bai_stone13}
Bai, X.-N., \& Stone, J. M. 2013,
\apj, 769, 76
\bibitem[Dauphas \& Pourmand(2011)]{dauphas_pourmand11}
Dauphas, N., \& Pourmand, A. 2011,
\nat, 473, 489
\bibitem[Dr\c{a}\.zkowska et al.(2016)]{drazkowska_etal16}
Dr\c{a}\.zkowska, J., Alibert, Y., \& Moore, B., 2016,
\aap, 591, A105
\bibitem[Gressel et al.(2015)]{gressel_etal15}
Gressel, O., Turner, N. J.; Nelson, R, P., \& McNally, C. P. 2015, \apj, 801, 84
\bibitem[Hansen(2009)]{hansen09}
Hansen, B. M. S. 2009,
\apj, 703, 1131
\bibitem[Hasegawa et al.(2017)]{hasegawa_etal17}
Hasegawa, Y., Okuzumi, S. Flock, M. \& Turner, N. J. 2017,
\apj, 845, 31
\bibitem[Hayashi(1981)]{hayashi81}
Hayashi, C. 1981,
Prog. Theor. Phys. Suppl., 70, 35
\bibitem[Kokubo \& Ida(2002)]{kokubo_ida02}
Kokubo, E., \& Ida, S. 2002,
\apj, 581, 666
\bibitem[Morbidelli et al.(2016)]{morbidelli_etal16}
Morbidelli, A., Bitsch, B., Crida, A., et al. 2016,
\icarus, 267, 368
\bibitem[Morishima et al.(2010)]{morishima_etal10}
Morishima, R., Stadel, J., \& Moore, B. 2010,
\icarus, 207, 517
\bibitem[Nagasawa et al.(2005)]{nagasawa_etal05}
Nagasawa, M., Lin, D. N. C., \& Thommes, E. 2005,
\apj, 635, 578
\bibitem[Ogihara et al.(2007)]{ogihara_etal07}
Ogihara, M., Ida, S., \& Morbidelli, A. 2007,
\icarus, 188, 522
\bibitem[Ogihara et al.(2015)]{ogihara_etal15}
Ogihara, M., Kobayashi, H., Inutsuka, S., \& Suzuki, T. K. 2015,
\aap, 579, A65
\bibitem[Ogihara et al.(2017)]{ogihara_etal17}
Ogihara, M., Kokubo, E., Suzuki, T. K., Morbidelli, A., \& Crida, A. 2017,
\aap, 608, A74
\bibitem[Ogihara et al.(2018b)]{ogihara_etal18b}
Ogihara, M., Kokubo, E., Suzuki, T. K., \& Morbidelli, A. 2018, 
\aap, in press. 
\bibitem[Oka et al.(2011)]{oka_etal11}
Oka, A., Nakamoto, T., \& Ida, S. 2011,
\apj, 738, 141
\bibitem[Paardekooper et al.(2010)]{paardekooper_etal10}
Paardekooper, S. -J., Baruteau, C., Crida, A. \& Kley, W. 2010,
\mnras, 401, 1950
\bibitem[Paardekooper et al.(2011)]{paardekooper_etal11}
Paardekooper, S. -J., Baruteau, C., \& Kley, W. 2011,
\mnras, 410, 293
\bibitem[Raymond et al.(2009)]{raymond_etal09}
Raymond, S., O'Brien D. P., Morbidelli, A., \& Kaib, N. A. 2009,
\icarus, 203, 644
\bibitem[Shakura \& Sunyaev(1973)]{shakura_sunyaev73}
Shakura, N. I., \& Sunyaev, R. A. 1973,
\aap, 24, 337
\bibitem[Simon et al.(2015)]{simon_etal15}
Simon, J. B., Lesur, G., Kunz, M. W., \& Armitage, P. J. 2015,
\mnras, 454, 1117
\bibitem[Simon et al.(2016)]{simon_etal16}
Simon, J. B., Armitage, P., Li, R., \& Youdin, A. N. 2016,
\apj, 822, 55
\bibitem[Suzuki \& Inutsuka(2009)]{suzuki_inutsuka09}
Suzuki, T. K., \& Inutsuka, S. 2009,
\apj, 691, L49
\bibitem[Suzuki et al.(2010)]{suzuki_etal10}
Suzuki, T. K., Muto, T., \& Inutsuka, S. 2010,
\apj, 718, 1289
\bibitem[Suzuki et al.(2016)]{suzuki_etal16}
Suzuki, T. K., Ogihara, M., Morbidelli, A., Crida, A. \& Guillot, T. 2016,
\aap, 596, A74
\bibitem[Walsh et al.(2011)]{walsh_etal11}
Walsh, K. J., Morbidelli, A., Raymond, S. N., O’Brien D. P., \& Mandell, A. M. 2011, 
\nat, 475, 206
\bibitem[Weidenschilling(1977)]{weidenschilling77}
Weidenschilling, S. J. 1977, Ap\&SS, 51, 153
\end{thebibliography}

\appendix
\section{Radial drift due to aerodynamical gas drag}
\label{sec:app1}

We use a force formula of aerodynamical gas drag on planetesimals developed by \citet{adachi_etal76};
\begin{eqnarray}
\mbox{\boldmath $F$}_{\rm aero} = - \frac{1}{2} C_{\rm D} \pi R^2 \rho_{\rm g} \Delta u \mbox{\boldmath $\Delta u$},
\end{eqnarray}
wehre $C_{\rm D}, \rho_{\rm g},$ and $\Delta u$ are the gas drag coefficient, the gas density, and the relative velocity of the planetesimal to the disk gas. The relative velocity is expressed by 
\begin{eqnarray}
\Delta u \simeq \left( \frac{5}{8} e^2 + \frac{1}{2} i^2 + \eta^2\right)^{1/2} v_{\rm K}.
\end{eqnarray}
Here we use the random velocity relative to the local Keplerian motion and the gas velocity ($v_{\rm gas} = [1 - \eta] v_{\rm K}$). The stopping time is given by $t_{\rm stop} = M \Delta u / |\mbox{\boldmath $F$}_{\rm aero}|$. Then the drift timescale is
\begin{eqnarray}
t_{\rm drift} &\simeq& \frac{t_{\rm stop}}{2 |\eta|} \label{eq:t_drift}\\
&\simeq& 2 \times 10^5  \left(\frac{\eta}{10^{-3}}\right)^{-1}
\left(\frac{C_{\rm D}}{2}\right)^{-1}  \left(\frac{M}{10^{16}~{\rm g}}\right)^{1/3}
\left(\frac{\rho}{3~{\rm g/cm^3}}\right)^{2/3} \nonumber \\ && \times
 \left(\frac{\Sigma_{\rm g}}{10^3~{\rm g/cm-^2}}\right)^{-1} 
 \left(\frac{H/r}{0.05}\right) \left(\frac{r}{1~{\rm au}}\right)^{3/2} \nonumber \\ && \times
\left(\frac{\sqrt{\frac{5}{8} e^2 + \frac{1}{2} i^2 + \eta^2}}{0.01}\right)^{-1} {\rm yr},
\end{eqnarray}
where $\rho$ is the internal density of planetesimals. In Section~\ref{sec:radial_drift}, $e=0.01$ and $i^2 = e^2 / 4$ are used. To obtain a more accurate orbital evolution of planetesimals, we need to calculate the evolution of $e$ and $i$, which should be considered in a future work.

\begin{figure}
\resizebox{1.0 \hsize}{!}{\includegraphics{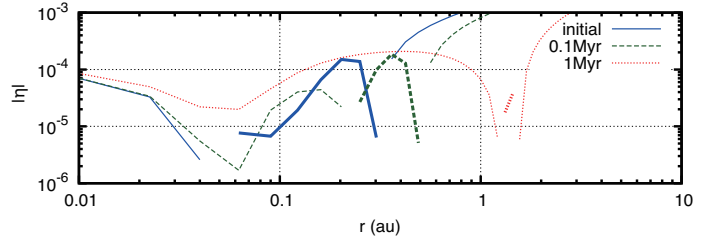}}
\caption{Time evolution of the absolute value of the scaling factor of pressure gradient force $\eta$ for an MRI-inactive disk based on \citet{suzuki_etal16}. Thick lines correspond to the region where $\eta$ is negative.}
\label{fig:app}
\end{figure}

Figure~\ref{fig:app} shows the time evolution of $\eta$ for an MRI-inactive disk. A similar plot is shown in Fig.~10 of \citet{suzuki_etal16}. The direction of radial drift is determined by the sign of $\eta$. 
Although Eq.~(\ref{eq:t_drift}) can be  a little more complicated for $|\eta|<e^2$ (see Eq.~(4.21) of \citealt{adachi_etal76}), outward drift is clearly seen for $\eta \lesssim -1 \times10^{-4}$.
Thick lines represent the region of negative $\eta$. There exists a region with positive pressure gradient in the inner region ($< 1~{\rm au}$), which move outwards in time. The positive pressure gradient is seen in region between 0.05 and 0.4 au at the initial time, while the region is between 0.2 and 0.6 au at $t= 0.1~{\rm Myr}$. Therefore, all planetesimals inside $\sim 1~{\rm au}$ can undergo outward radial drift before $t = 1~{\rm Myr}$ in results of Section~\ref{sec:radial_drift}.

\end{document}